\documentstyle[12pt,psfig]{article}
\let\section=\subsection     \let\subsection=\subsubsection                %%
\newcommand{\gc}{\gamma}
\newcommand{\gd}{\delta}

\begin{document}
\renewcommand{\thefootnote}{\fnsymbol{footnote}}
\begin{center}
   {\large \bf DEUTERON FORMATION IN NUCLEAR MATTER}\\[2mm]
   {\large \bf  WITHIN THE FADDEEV  APPROACH\footnote{Work 
       supported by the Deutsche
       Forschungsgemeinschaft. }\footnote{Talk presented by M. Beyer.}}\\[5mm] 
   M.~BEYER, C.~KUHRTS, G.~R\"OPKE \\[5mm]
   {\small \it  FB Physik, Universit\"at Rostock, 18051 Rostock,
     Germany \\[8mm] } 
\end{center}

\begin{abstract}\noindent
  We consider deuteron formation in heavy ion collisions at
  intermediate energies. The elementary reaction rates ($Nd
  \rightarrow NNN$ etc.) in this context are calculated using rigorous
  Faddeev methods. To this end an in-medium Faddeev equation that
  consistently includes the energy shift and Pauli blocking effects
  has been derived and solved numerically. As a first application we
  have calculated the life-time of deuteron fluctuations for nuclear
  densities and temperatures typical for the final stage of heavy ion
  collisions. We find substantial differences between using the
  isolated and the in-medium rates.
\end{abstract}

\section{Introduction}
Nuclear matter is an example of a strongly correlated many particle
system. One prominent consequence is the formation of bound states
(clusters, fragments) observed in heavy ion reactions. Here we address
the formation of deuterons at intermediate energies, i.e. for $E/A\leq
200$ MeV/u.  Within the quantum statistical approach to describe the
complicated dynamics we employ the Green function method~\cite{kad62}.
The cluster mean field approximation~\cite{roepke} decouples the
hierarchy and leads to rigourous few-body equations for the two-,
three-, four-particle correlations.  This method is tiedly connected
to the self consistent RPA approach extended to finite
temperatures~\cite{schuck}.

Deuteron formation is directly related to the $Nd\rightarrow NNN$
break-up cross section. Photoinduced reactions have also been
considered~\cite{pawel}.  Because of the energies considered, pion
induced reactions can be neglected.

Treating deuteron formation within the cluster Hartree-Fock
approximation allows us to consistently include all medium
modifications as they appear in the respective two- and three-body
equations. These are the self energy and the Pauli blocking effects.

\section{Reactions}

The quantity of interest in the quantum statistical approach is the
generalized quantum Boltzmann equation for the nucleon $f_N$, deuteron
$f_d$, etc. distributions. Here, we consider the collision integral to
show the relevance of three-body reaction rates. The feeding of the
nucleon density is driven by the collision integral (see e.g.
Ref.~\cite{dan91} for an application to heavy ion collisions)
\begin{equation}
{\cal J}_N(p,t)={\cal J}^{\rm out}_N(p,t) f_N(p,t)
-{\cal J}^{\rm in}_N(p,t) \bar f_N(p,t),
\end{equation}
where we have used $\bar f_N=1- f_N$. To be more explicit we give
${\cal J}^{\rm out}_N(p,t)$,
\begin{eqnarray}
{\cal J}^{\rm out}_N(p,t)&=&
\int d^3k\int d^3k_1d^3k_2\; |\langle kp|T|k_1k_2\rangle|^2\;
\bar f_N(k_1,t)\bar f_N(k_2,t) f_N(k,t)\nonumber\\
&&+\int d^3k\int d^3k_1d^3k_2d^3k_3\;
 |\langle kp|U_0|k_1k_2k_3\rangle|^2\nonumber\\&&
\qquad\times
\bar f_N(k_1,t)\bar f_N(k_2,t)\bar f_N(k_3,t)f_d(k,t)\nonumber\\&&
+\dots\label{eqn:react}
\end{eqnarray}
where dots stand for other possible contributions mentioned before.
The quantity $U_0$ appearing in (\ref{eqn:react}) is the break-up
transition operator for $Nd\rightarrow NNN$. For the isolated
three-body problem $U_0$ determines the break-up cross section
$\sigma_0$ via
\begin{eqnarray}
\sigma_0 &=& \frac{1}{|v_d-v_N|} \frac{1}{3!}
\int d^3k_1d^3k_2d^3k_3\;
 |\langle kp|U_0|k_1k_2k_3\rangle|^2\nonumber\\
&&\qquad\qquad\times2\pi\delta(E_f-E_i)\;(2\pi)^3\delta^{(3)}(k_1+k_2+k_3).
\end{eqnarray}
So far the strategy has been to implement the {\em experimental} cross
section into the above equation. This has then been solved, for a
specific heavy ion collision~\cite{dan91}.  Using experimental cross
sections respectively isolated cross sections may not be sufficient in
particular in the lower energy regime. The cross section itself
depends on the medium, e.g. blocking of internal lines or self energy
corrections of the respective three-body Green functions. To this end
we have derived a three-body Faddeev type equation~\cite{bey96,bey97}.
We use the AGS formalism~\cite{AGS} for the three-body algebra and
solve the respective equations numerically.  The equation for the
three-particle Green function derived within the cluster Hartree-Fock
approximation reads
\begin{equation}
G_3(z) = \frac{\bar f_1\bar f_2\bar f_3+f_1f_2f_3}
{z-\varepsilon_1-\varepsilon_2-\varepsilon_3}
+\frac{(\bar f_1\bar f_2 -f_1f_2) V(12) + {\rm cycl. perm.}}
{z-\varepsilon_1-\varepsilon_2-\varepsilon_3}\; G_3(z),
\end{equation}
where $\varepsilon$ denotes the quasi particle energies evaluated in
Hartree-Fock approximation and $f_1=f(\varepsilon_1)$ the Fermi
functions $f(\varepsilon)=(\exp[\beta(\varepsilon-\mu)+1]^{-1}$ with
the inverse temperature $\beta$ and the chemical potential $\mu$ (for
the time being we assume symmetric nuclear matter). Here, we use
equilibrium distributions to  solve Faddeev equations. This
is justified within the linear response theory, where
nonequilibrium quantities are expressed through equilibrium ones,
because of small fluctuations only. The question of self consistency
has been addressed for the much simpler case of two-particle
correlations, e.g. in Ref.~\cite{schnell}.

Within the AGS formalism (extended here to finite temperatures and
densities) the break-up operator $U_0$ is simply related to the
elastic/rearrangement scattering amplitude $U_{\alpha\beta}$
connecting the channels\footnote{The channel notation
  $\alpha,\beta=1,2,3$ labels the respective spectator in the
  three-body system.}  $\beta\rightarrow\alpha$. It is therefore
sufficient to present the AGS equation for the transition operator
$U_{\alpha\beta}$ only, viz.
\begin{equation}
U_{\alpha\beta}(z)= \bar\delta_{\alpha\beta}
\left[ \frac{\bar f_1\bar f_2\bar f_3+f_1f_2f_3}
{z-\varepsilon_1-\varepsilon_2-\varepsilon_3}\right]^{-1}
+\sum_{\alpha\neq\gamma=1}^3
T_3^{(\gamma)} \frac{\bar f_1\bar f_2\bar f_3+f_1f_2f_3}
{z-\varepsilon_1-\varepsilon_2-\varepsilon_3}\; U_{\gamma\beta}(z)
\end{equation}
with $\bar\delta=1-\delta$ and $\bar f=1-f$. The two-body $t$ matrix
$T_3^{(\gc)}$ has been solved on the same footing consistently
including all medium effects, i.e.  $T_3^{(\gc)}$ is the solution of
the in-medium two-body problem, e.g. for $\gc=3$
\begin{equation}
T_3^{(3)}= (1-f_3+g(\varepsilon_1+\varepsilon_2))^{-1}\;
V_2+V_2\;\frac{\bar f_1\bar f_2 - f_1f_2}
{z - \varepsilon_1 - \varepsilon_2- \varepsilon_3}\;T_3^{(3)},
\end{equation}
where $g(\omega)=(\exp[\beta(\omega-2\mu)-1]^{-1}$ is the Bose
function for two nucleons.
\section{Results}
\begin{figure}[t]
\begin{minipage}{0.48\textwidth}
\psfig{figure=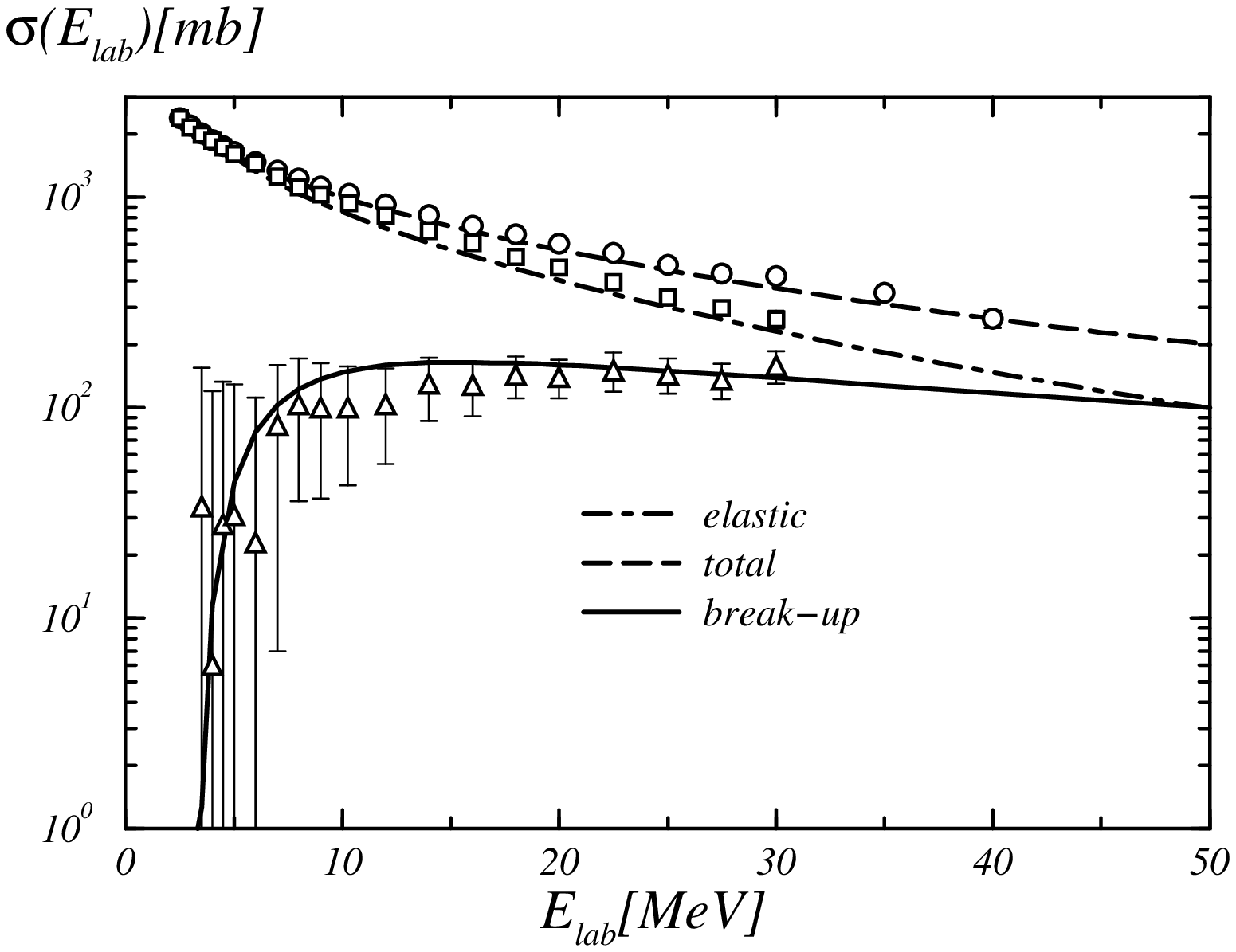,width=\textwidth}
\caption{\label{fig:isocross}  A comparison of the total, elastic,  
  and break-up cross sections $nd\rightarrow nd$, $nd\rightarrow
  nnp$ with the experimental data of
  Ref.~\protect{\cite{sch83}}.}
\end{minipage}
\hfill
\begin{minipage}{0.48\textwidth}
\psfig{figure=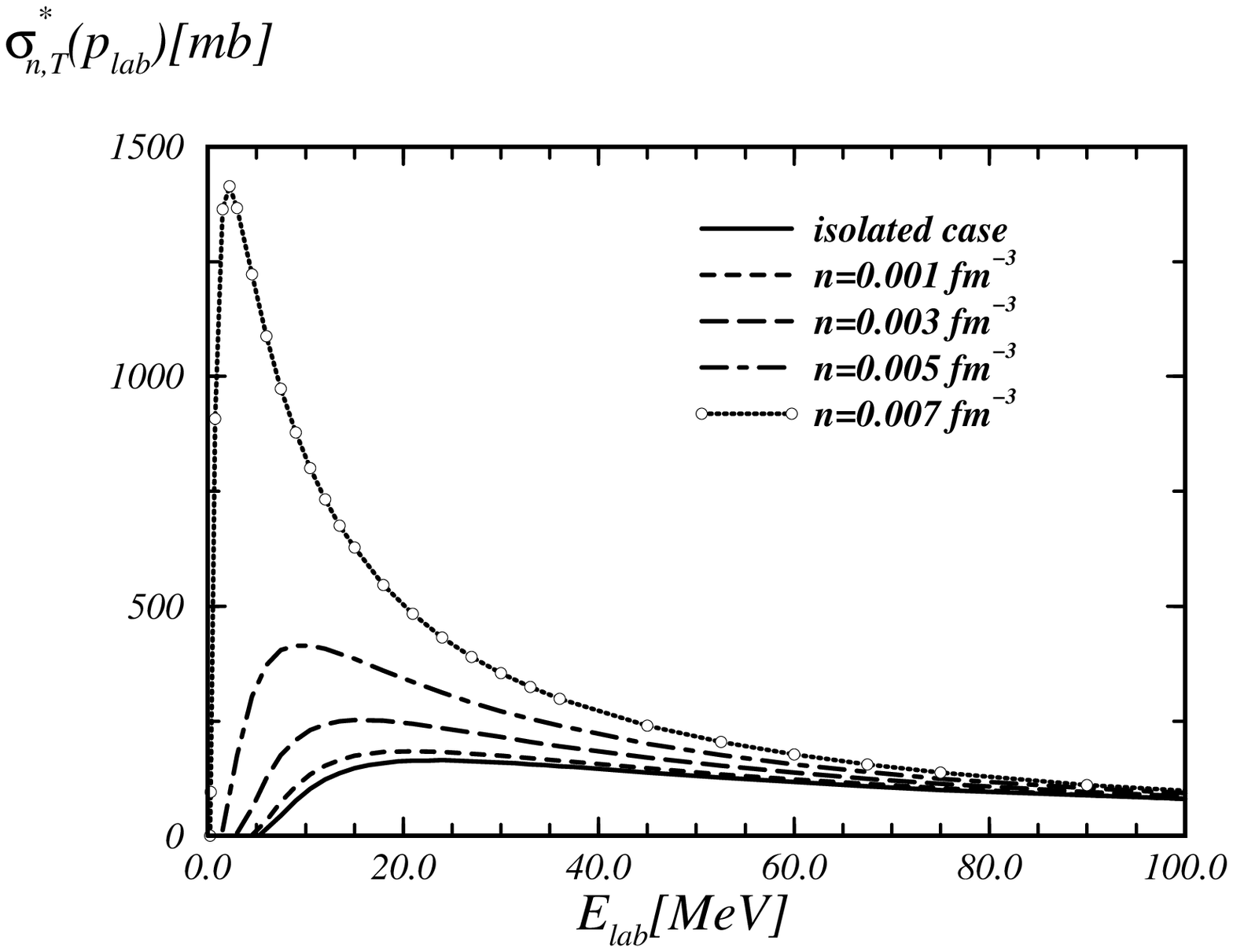,width=\textwidth}
\caption{\label{fig:cross} In-medium break-up cross section at $T=10$
  MeV. Isolated cross section is shown as solid line, other lines  
  show different nuclear densities.  }
\end{minipage}
\end{figure}

The AGS equations have been solved for a separable Yamaguchi
potential. 
%Since the three body dynamics if driven by the two-body $t$
%matrix, that at this energies is dominated by the deuteron pole and
%the $^1$S$_0$ antibound state, this approximation leads to a
%reasonable description of the cross section. 
To get an impression of
the quality of the calculation the isolated cross section is given in
Fig.~\ref{fig:isocross} along with the experimental data on neutron
deuteron scattering~\cite{sch83}.

From inspection of Fig.~\ref{fig:cross} we see that the in-medium cross 
section is significantly enhanced compared to the isolated on. The
threshold is shifted to smaller energies, which is because the binding 
energy of the deuteron becomes smaller. We observe that for higher
energies the medium dependence of the cross section becomes much
weaker, which {\em a posteriori} justifies the use of isolated cross sections 
(along with the impulse approximation) when higher energies are
considered~\cite{dan91}. 

\begin{figure}[tbh]
\begin{minipage}[t]{0.48\textwidth}
\psfig{figure=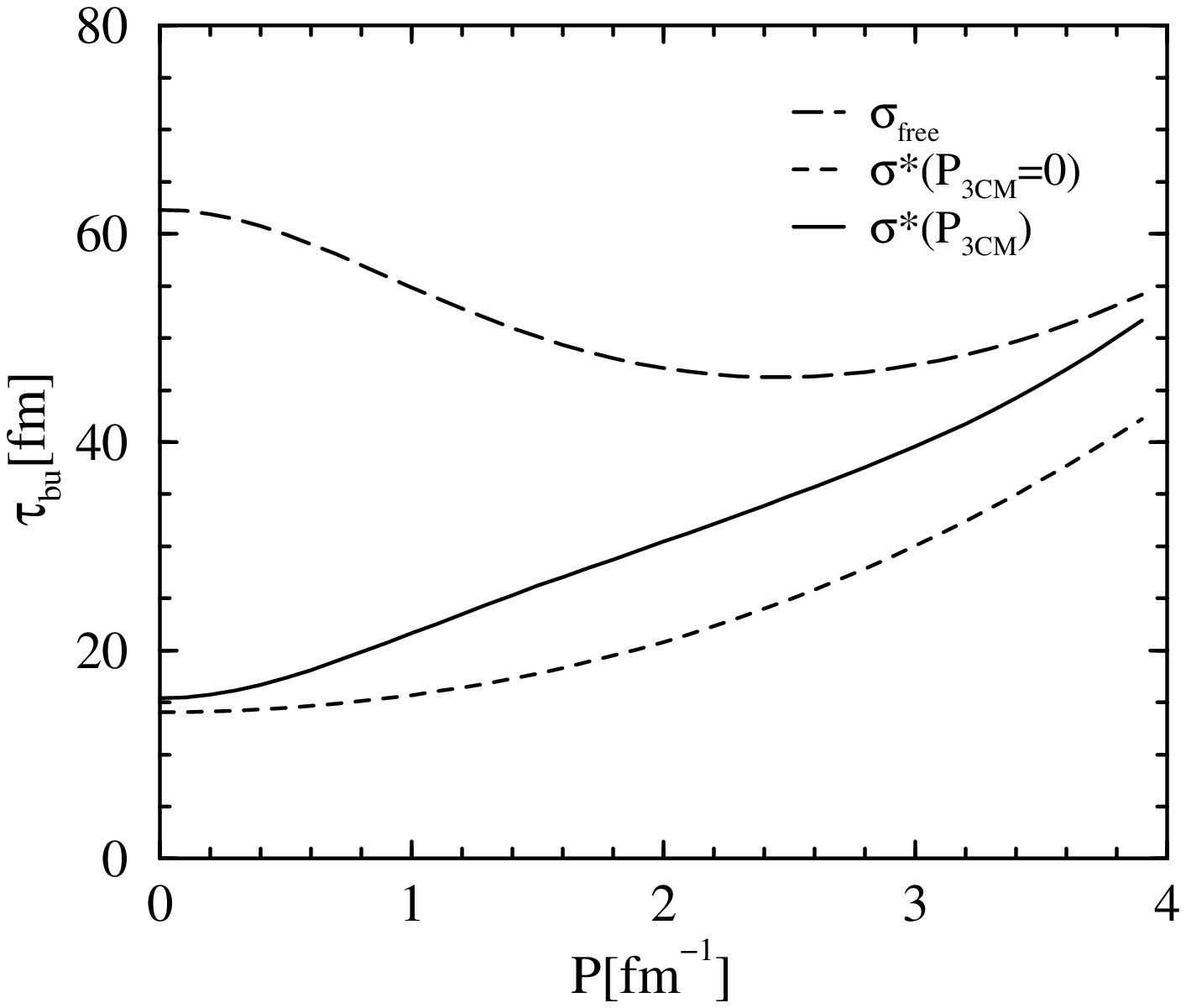,width=0.95\textwidth}
\caption{\label{fig:life} Deuteron break-up time at $T=10$ MeV and  
  nuclear density $n=0.007$ fm$^{-3}$. Solid line with medium
  dependent cross section a given in Fig.~\protect\ref{fig:cross}.
  Short dashed with $P_{\rm cm}=0$ and dashed line isolated cross
  section. } 
\end{minipage}
\hfill
\begin{minipage}[t]{0.48\textwidth}
\psfig{figure=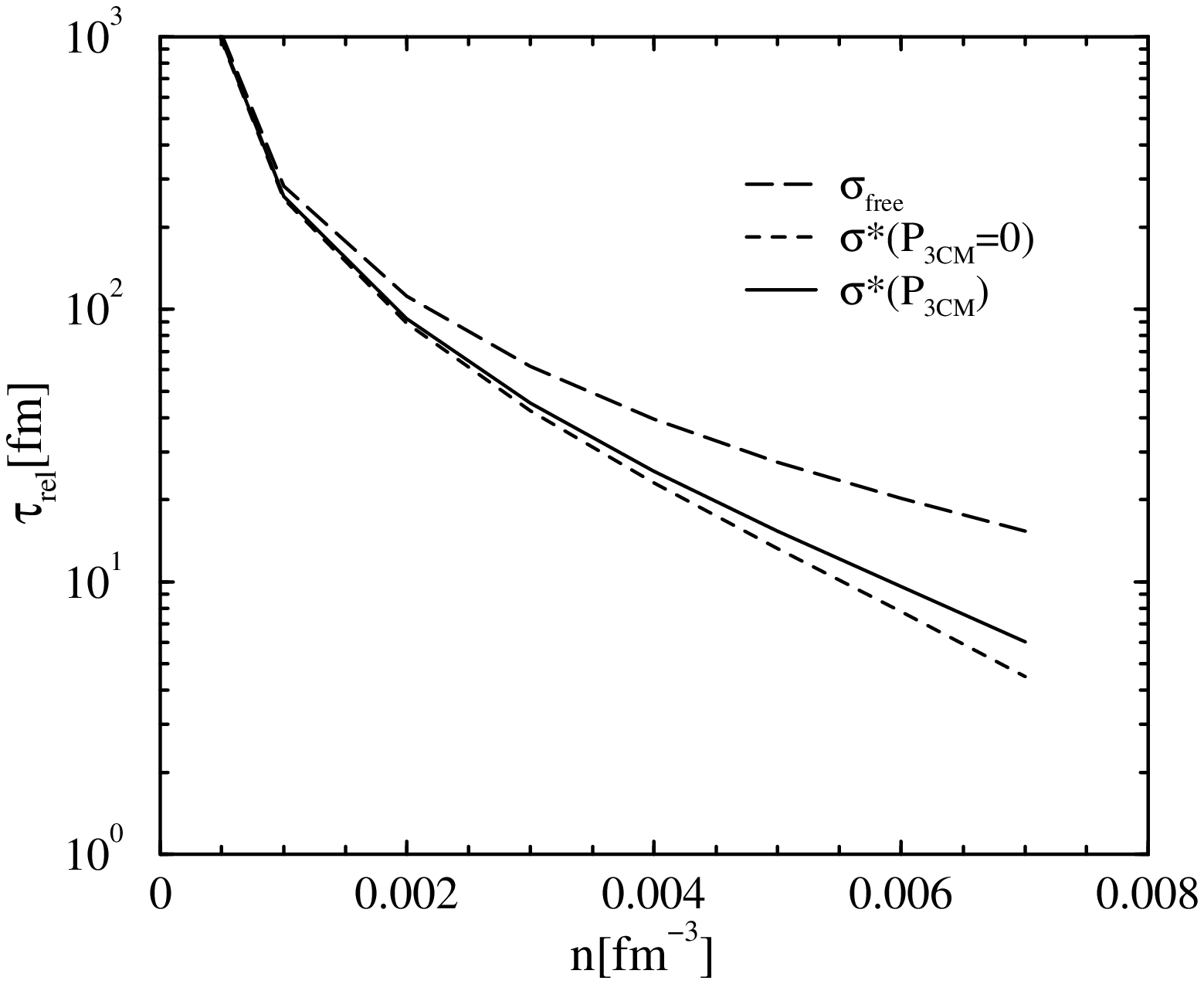,width=\textwidth}
\caption{\label{fig:Trel} Relaxation time for small fluctuations of
  the deuteron density from chemical equilibrium at a temperature of
  $T=10$ MeV. Line coding as in Fig.~\protect\ref{fig:life}}
\end{minipage}
\end{figure}

From linearizing the Boltzmann equation it is possible to define a
break-up time for small fluctuations of the deuteron distributions. For
small fluctuations $\gd f(t)=f_d(t)-f_d^0$ from the equilibrium
distribution $f_d^0$ linear response leads to
\begin{equation}
\partial_t \gd f_d(P,t) = -\frac{1}{\tau_{\rm bu}}
\gd f_d(P,t)
\end{equation}
where the ``life time'' of deuteron fluctuations has been
introduced~\cite{bey97},
\begin{equation}
\tau^{-1}_{\rm bu} 
= \frac{4}{3!} \int dk_Nd^3k_1 d^3k_2 d^3k_3\;
\left|\langle kp|U_0|k_1k_2k_3\rangle\right|^2
%{\rm Tr}({\cal M}_0\rho_i{\cal M}_0^\dagger)
\; \bar f_{1}\bar f_{2}\bar f_{3}f(k_N)
\;2\pi\gd(E - E_0).
\label{eqn:lifetime}
\end{equation}
which can be related to the break-up cross section given in
Fig.~\ref{fig:cross}. For low densities the life time (as a function of
the deuteron momentum $P$) and the inverse life time, i.e. the width,
at $P=0$ along with the deuteron binding energy for comparison is
shown in Fig.~\ref{fig:life}. These times have to be
compared to the approximate duration of the heavy ion collision of
about 200 fm.

Another important time scale is the chemical relaxation time for small
fluctuations of the deuteron density $\delta n_d(t)=n_d(t)-n^0_d$ from
the equilibrium distribution $n^0_d$.  Using detailed balance and
linearized rate equations the relaxation time is given through
\begin{equation}
\frac{d}{dt} \gd n_d(t) = -\frac{1}{\tau_{\rm rel}}
\gd n_d(t).
\end{equation}
The basic quantity driving the time scale is again the break-up cross
section
\begin{equation}
\frac{1}{\tau_{\rm rel}} = 
\int d^3p_N\;d^3p_d\;f(k_N)g(p_d) |v_d-v_N|\; \sigma_0(E)
\;\frac{n^0_N+4n^0_d}{n^0_Nn^0_d}
\end{equation}
The resulting relaxation time as a function of the uncorrelated
nuclear density is given in Fig.~\ref{fig:Trel}.

\section{Conclusion and Outlook}

Our results show that medium dependent cross sections in the
respective collision integrals lead to {\em shorter reaction time
  scales}.  Chemical processes become faster. This also effects the
elastic rates that are related to thermal equilibration.

The basis of this result is the cluster Hartree-Fock approach that in
our approximation includes correlations up to three particles in a
consistent way. The equations driving the correlations are rigorous.
The respective one-, two- and three-body equations are solved, in
particular for the three-particle case Faddeev/AGS type equations
have been derived in Ref.~\cite{bey96,bey97}. The AGS approach is
particularly appealing since it allows generalizations to $n$-particle
equations in a straight forward way. Results for the three-body bound
states in medium will be published elsewhere~\cite{triton}. As
expected form the deuteron case, the triton binding energy changes
with increasing density up to the Mott density, where $E_t=0$.

The production rates, spectra etc.  of light charged particles in
heavy ion collisions at intermediate energies may change because of
the much smaller time scales induced through the medium dependence
compared to the use of free cross sections (respectively experimental
cross sections). To this end some notion of the relevant densities and 
temperatures during the heavy ion collision (during the final stage)
should be achieved.

\end{document}